\documentclass[aps,prl,twocolumn,showpacs,floatfix,superscriptaddress]{revtex4-1}
\usepackage{graphicx,color}
\usepackage{amsfonts}
\usepackage[figuresright]{rotating}
\usepackage{amssymb}
\usepackage{amsmath}
\usepackage{psfrag}
\usepackage{subfigure}
\usepackage{multirow}
\usepackage{tabularx}
\usepackage{textcomp}
\usepackage{units}
\usepackage{hyperref}
\hypersetup{
 pdfnewwindow=true, colorlinks=true,
 linkcolor=blue, anchorcolor=blue,
 citecolor=blue, filecolor=blue,
 menucolor=blue, urlcolor=blue}

\usepackage{graphicx}
\usepackage{dcolumn}
\usepackage{bm}
\usepackage{color}

\makeatletter

\begin{document}
\title{Photostrictive two-dimensional materials in the monochalcogenide family}

\author{Raad\ \surname{Haleoot}}
\affiliation{Department of Physics, University of Arkansas, Fayetteville, AR 72701, USA}
\affiliation{Department of Physics at the College of Education, University of Mustansiriyah, Baghdad, Iraq}

\author{Charles\ \surname{Paillard}}
\affiliation{Department of Physics, University of Arkansas, Fayetteville, AR 72701, USA}
\affiliation{Institute for Nanoscience and Engineering, University of Arkansas, Fayetteville, AR 72701, USA}

\author{Mehrshad\ \surname{Mehboudi}}
\affiliation{Department of Physics, University of Arkansas, Fayetteville, AR 72701, USA}

\author{Bin\ \surname{Xu}}
\affiliation{Department of Physics, University of Arkansas, Fayetteville, AR 72701, USA}
\affiliation{Institute for Nanoscience and Engineering, University of Arkansas, Fayetteville, AR 72701, USA}

\author{L.\ \surname{Bellaiche}}
\affiliation{Department of Physics, University of Arkansas, Fayetteville, AR 72701, USA}
\affiliation{Institute for Nanoscience and Engineering, University of Arkansas, Fayetteville, AR 72701, USA}

\author{Salvador\ \surname{Barraza-Lopez}}
\email{sbarraza@uark.edu}
\affiliation{Department of Physics, University of Arkansas, Fayetteville, AR 72701, USA}

\begin{abstract}
Photostriction is predicted for SnS and SnSe monolayers, two-dimensional ferroelectrics with rectangular unit cells (the lattice vector $\mathbf{a}_1$ is larger than $\mathbf{a}_2$) and an intrinsic dipole moment parallel to $\mathbf{a}_1$. Photostriction in these two-dimensional materials is found to be related to the structural change induced by a screened electric polarization in the photoexcited electronic state (i.e., a converse piezoelectric effect) that leads to a compression of $a_1$ and a comparatively smaller increase of $a_2$ for a reduced unit cell area.  The structural change documented here is ten times larger than that observed in BiFeO$_3$, making monochalcogenide monolayers an ultimate platform for this effect. This structural modification should be observable under experimentally feasible densities of photexcited carriers on samples that have been grown already, having a potential usefulness for light-induced, remote mechano-opto-electronic applications.
\end{abstract}

\date{\today}

\maketitle

A truly novel opto-mechanical coupling in two-dimensional (2D) ferroelectric materials awaits to be discovered. Photostriction --the creation of non-thermal strain upon illumination \cite{sixties,sixties2,kundys2015photostrictive,review16}-- has been well-documented in three-dimensional ferroelectrics such as SbSI \cite{Uchino} and BiFeO$_3$ \cite{ps1,ps2}. It has been suggested to be driven by the large voltage build-up caused by a photovoltaic effect and the resulting converse piezoelectricity \cite{Fridkin}, and it may be useful for applications such as  remotely-switchable memory devices \cite{memory} and light-induced actuators \cite{Laurent}. The earliest studied photostrictive material, SbSI, transitions from a ferroelectric onto a paraelectric at a critical temperature $T_c<300$ K. As photostrictive effects are larger in the ferroelectric phase, $T_c$ can be increased above 300 K on SbSI ceramics which have smaller domain sizes and display a non-uniform stoichiometry nevertheless. The photostriction response time increases with sample thickness due to a reduced penetration depth, being a few seconds on bulk samples \cite{Uchino}. On the other hand, ferroelectric films show photostriction within a few picoseconds \cite{ref39,ps0,ps_1,ps3,fast1}, and even when the photoexcited electron-hole pair is localized \cite{localized,localized2}.

The growing interest on the interactions of light with 2D materials \cite{Castellanos,Yang,Fogler1,Fogler2} makes a study of illumination leading to non-trivial structural deformations an interesting and timely endeavor. As long as a photoexcited state induces some amount of charge redistribution --which is a quite reasonable physical assumption-- any material is expected to change shape as the structure is let to relieve the stress induced by the photoexcited carriers. Here, the surprising result is the {\em rather large magnitude of such structural change for 2D ferroelectrics}, that originates from an inverse piezoelectric effect upon illumination.

Two-dimensional ferroelectrics in the group IV monochalcogenide family (GeS, GeSe, SnS, SnSe, among others) \cite{Kaxiras,Hennig,Tomanek,Mehboudi2016,Chang274,cheng,Mehboudi2016b,Carvalho,Yang2,unpublished} undergo a ferroelectric-to-paraelectric transition with a transition temperature that is tunable by atomic number \cite{Mehboudi2016}. Following the numerical approach proven to capture photostriction of BiFeO$_3$ \cite{Laurent}, the photostriction of SnS and SnSe is successfully predicted hereby. This effect could be readily observed in recent experimental setups as the one in Ref.~\cite{Chang274}.

The challenge at hand and the computational approach are described first. Then, the two direct optical transitions to be employed to demonstrate the effect are motivated, and the anisotropic change of lattice parameters (photostriction) upon photoexcitation is documented. The decrease of the dipole moment and unit cell area seen in our numerical results are explained in terms of a photoinduced inverse piezoelectric effect and electronic pressure afterwards.

The concept is straightforward: one creates the effect of a direct optical transition at the valence and conduction band edges, {\em allowing the structure to relax the forces created in the photoexcited state}. In practice, capturing the effect requires approximations: specifically, the accuracy in forces needed to observe photostriction makes the Bethe-Salpeter approach \cite{Hedin,Louie} --the technique of choice for optical excitations in materials of reduced dimensionality-- prohibitively expensive, and the same could be said of a time-dependent approach to the problem \cite{Ullrich}. Indeed, photostriction under a density of photoexcited carriers $n_c$ changes lattice parameters $|\Delta a_i/a_{i,0}|\equiv |(a_i(n_c>0)-a_{i,0}|/a_{i,0}$ ($i=1,3$) to within $10^{-5}-10^{-4}$  in bulk samples \cite{kundys2015photostrictive}, making for a prohibitively expensive optimization of the electron-hole-pair hosting structure ($a_{i,0}\equiv a_i(n_c=0)$ here).

\begin{figure*}[tb]
\includegraphics[width=1.0\textwidth]{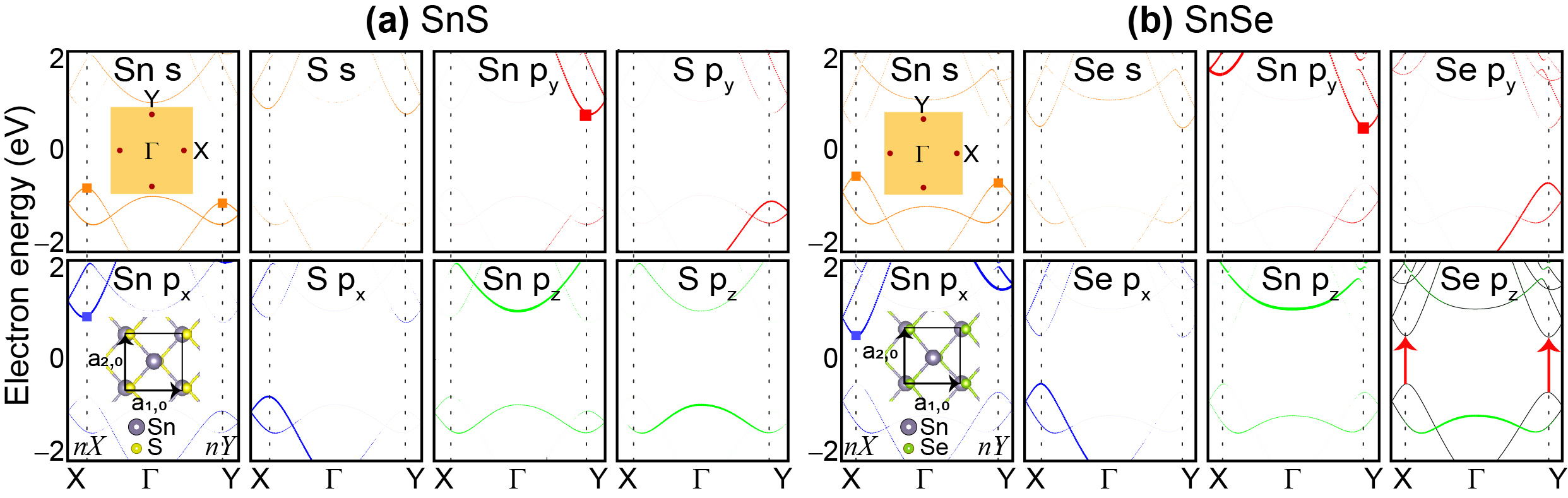}
\caption{Orbital-resolved electronic structure of (a) SnS and (b) SnSe monolayers, with probability related to the observed line thickness. Direct optical transitions among valence band edge states with $s$ symmetry (highlighted by orange squares) and conduction band edge states with $p_x$ ($p_y$) character at the $nX$ ($nY$) band edges are allowed by symmetry (shown by dotted vertical lines and highlighted by either blue or red squares in bandstructure plots). First Brillouin zones displaying the $\pm nX$ and $\pm nY$ band edges as dots, as well as structural unit cells, are shown as insets. The lower rightmost (Se $p_z$) subplot also includes a full bandstructure in black, and the two-direct optical transitions shown by red arrows as guides to the eye. }\label{fig:fig1}
\end{figure*}

However, the recent discovery of ferroelectricity in monochalcogenide monolayers \cite{Mehboudi2016,Chang274} gives an opportunity to extend this well-known effect into 2D  materials, and the structural deformation in photoexcited SnS and SnSe monolayers will be demonstrated using the same numerical technique \cite{Laurent} that successfully reproduces the experimentally observed photostriction of BiFeO$_3$ \cite{expt}.

G\"orling formulated the interacting, photoexcited Hamiltonian as a model non-interacting DFT Hamiltonian \cite{Gorling}, and the $\Delta$-self-consistent-field ($\Delta$SCF) method is a realization of G\"orling's approach that assumes a one-to-one correspondence between the excited states of a Kohn-Sham Hamiltonian and the real system \cite{Martin}. It creates a population imbalance akin to that produced from illumination, by depleting a finite number of electrons in the valence band and promoting them onto higher energy bands. $\Delta$SCF calculations of excited states for systems with reduced dimensions abound (e.g., Refs.~\cite{paper0,paper1,paper2}), and the $\Delta$SCF method as implemented in the {\em ABINIT} code \cite{gonze2009abinit} is employed to predict structural effects of direct optical transitions at the valley edges of ferroelectric SnS and SnSe monolayers here. Calculations were performed with GBRV projected-augmented-wave \cite{PAW} pseudopotentials \cite{GBRV} of the PBE type \cite{PBE}, which are known to underestimate the electronic band gap. Nevertheless, additional corrections make it prohibitive to demonstrate the effect within computational constraints.

Figure~\ref{fig:fig1} shows the electronic structure decomposed in states with $s$, $p_x$, $p_y$ or $p_z$ orbital symmetry and belonging to a specific atomic species (Sn or S in Fig.~\ref{fig:fig1}(a), and Sn or Se in Fig.~\ref{fig:fig1}(b)). Line thicknesses reflect the relative probability of finding a given orbital symmetry for a given band and chemical element. The lattice parameters for the ground-state structure shown as an inset in Fig.~\ref{fig:fig1}(a) (prior to photoexcitation; i.e., $n_c=0$) are $a_{1,0}=4.3087$ \AA{} and $a_{2,0}=4.0786$ \AA{} for the SnS monolayer. For the SnSe monolayer (inset in Fig.~\ref{fig:fig1}(b)) $a_{1,0}=4.4038$ \AA{} and $a_{2,0}=4.2918$ \AA. The unit cell area is $A_0=a_{1,0}a_{2,0}$.

Optical transitions require non-zero matrix elements $\langle p_{i,c}| \mathbf{r}|s_{v}\rangle$ for  wavefunctions with $|p_i\rangle$ symmetry in the conduction band and $|s\rangle$ symmetry in the valence band ($i=x$, $y$, $z$). In SnS, there are two main valleys near the corners of the first Brillouin zone that are located at 0.390$\mathbf{b}_1$ and 0.415$\mathbf{b}_2$ (vertical dotted lines in Fig.~\ref{fig:fig1}(a)) and labeled $nX$ and $nY$ as shorthand for $k-$points located ``near the $X-$high-symmetry point'' and ``near the $Y-$high-symmetry point,'' respectively ($\mathbf{b}_1$ and $\mathbf{b}_2$ are reciprocal lattice vectors). These valleys are located at 0.415$\mathbf{b}_1$ and 0.415$\mathbf{b}_2$ for SnSe (Fig.~\ref{fig:fig1}(b)).

According to Fig.~\ref{fig:fig1}, the group-IV element (Sn) has a larger probability to have an orbital $s-$symmetry at the $nX$ and $nY$ valence band edges for both SnS and SnSe monolayers (as emphasized by orange rectangles at such band edges). Similarly, a large probability is carried by Sn orbitals with $p_x$ ($p_y$) symmetry at the $nX$ ($nY$) conduction valley edge (red rectangles). This way, the non-zero $\langle p_{i,c}| \mathbf{r}|s_{v}\rangle$ matrix element originates from a Sn {\em intra-atomic} direct optical transition with linearly-polarized absorption band edges \cite{Mehboudi2016b,excitons}.

Illumination by pulsed laser sources can generate photoexcited carrier density fluences as high as $10^{13}-10^{14}$/cm$^2$ on MoS$_2$ samples \cite{Gedik}. It will be shown that a much smaller density is needed for the effect being presently described to be experimentally achievable, after discussing the $k-$point mesh employed in calculations.

\begin{figure}[tb]
\includegraphics[width=0.485\textwidth]{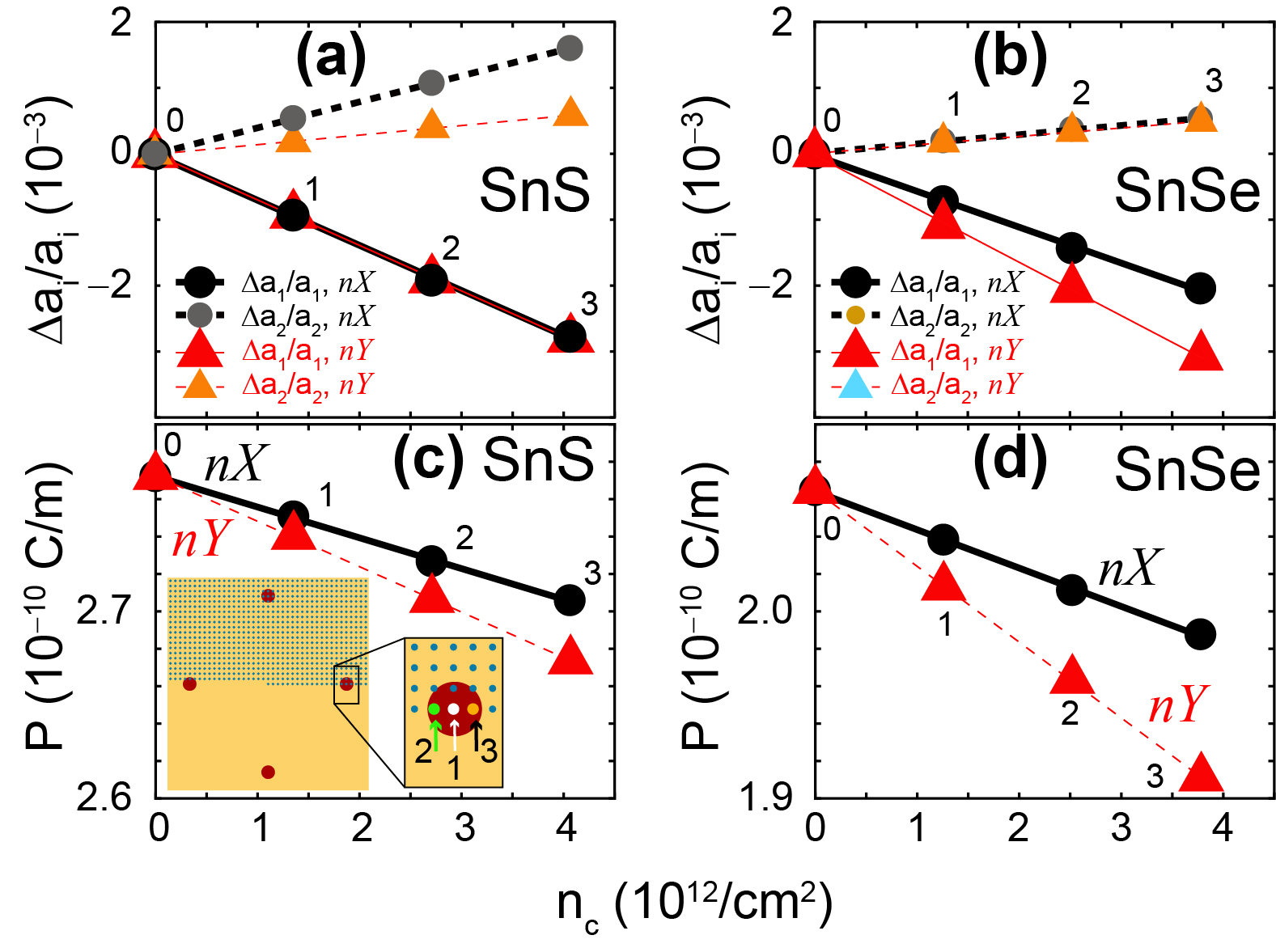}
\caption{Photostriction is the non-thermal change of lattice parameters upon irradiation by light demonstrated for (a) SnS and (b) SnSe monolayers. The change on $a_1$ and $a_2$ is one order of magnitude larger than that documented for BiFeO$_3$, for experimentally accessible excited carrier densities. (c)-(d) Photostriction leads to a decrease of the dipole moment $P$. The inset in subplot (c) shows the $k-$point mesh employed in calculations (the remainder of the first-Brilloin zone is included in calculations by symmetry), and the zoom-in exemplifies the three $k-$points ($n=1,2,3$) that were sequentially photoexcited about the $nX$ point.}\label{fig:fig2}
\end{figure}

Considering spin-orbit coupling (SOC), a regular 2D mesh containing $n_k^2$ equally-weighted $k-$points yields a density of $n_c=1/(n_k^2A_0)$ charge carriers per band per $k-$point per unit cell. The $k-$point mesh with $n_k=41$ --shown as an inset in Fig.~\ref{fig:fig2}(c)-- permits creating $n_c(n)=2^2n/(41^2A_0)\simeq 1.3n\times 10^{12}$/cm$^2$ excited charge carriers per band per unit cell. Here, the factor of four is due to the symmetry of the $k-$point mesh shown at the inset, and because carriers from two bands immediately below the bandgap are excited into two bands right above the bandgap that are slightly split due to SOC; the dependence of $n_c$ on $n=0,1,2$ or $3$ allows for a gradual increment of photoexcited carriers. Recalling that photostrition of bulk samples results on $|\Delta a_i|/a_{i,0}\simeq 10^{-4}-10^{-5}$ \cite{kundys2015photostrictive,expt,Laurent}, a demanding relaxation limit for structural forces of $5\times 10^{-8}$ Ha/Bohr and an energy cutoff of 40 Ha were employed in our calculations.

\begin{figure}[tb]
\includegraphics[width=0.485\textwidth]{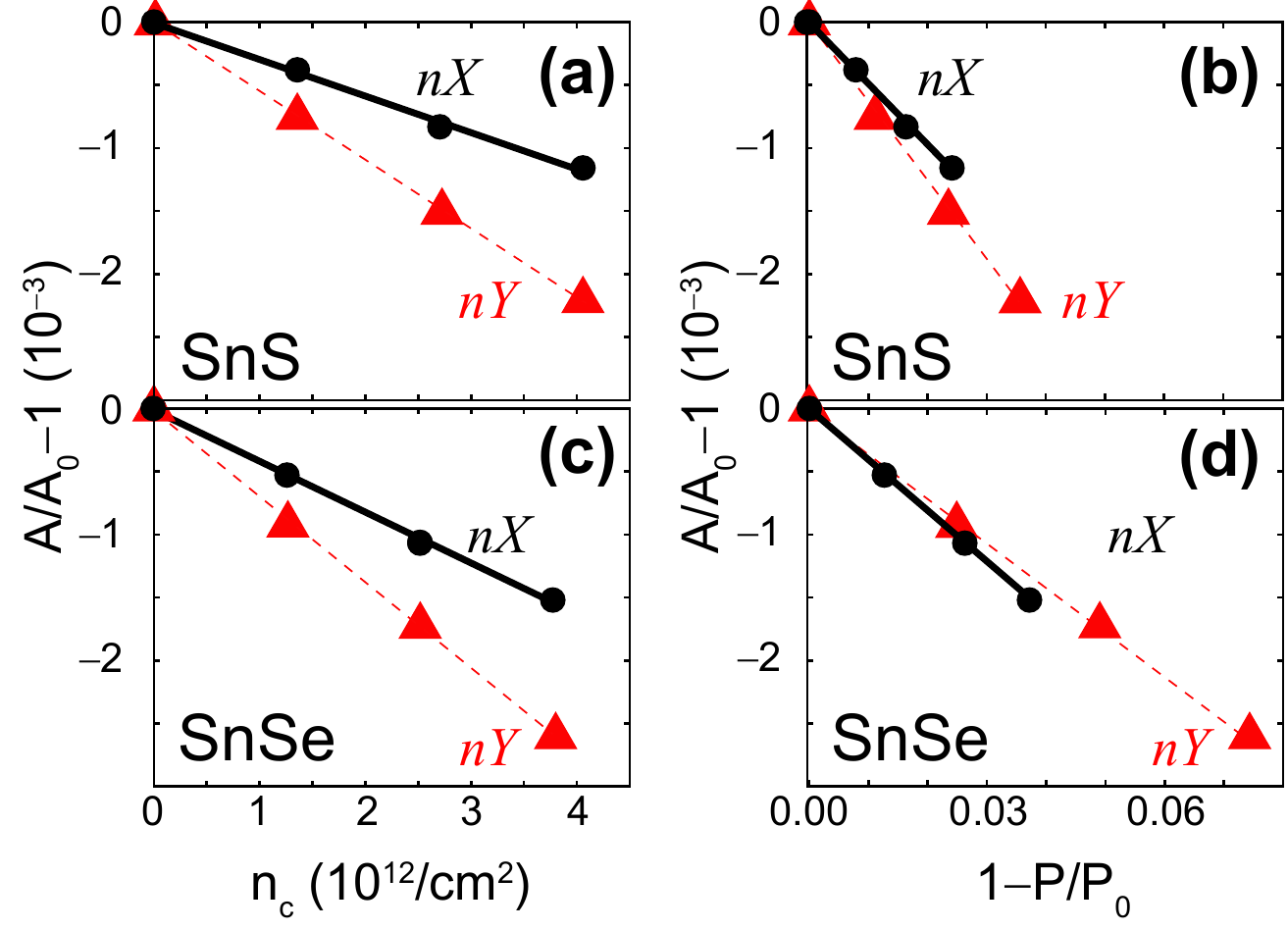}
\caption{(a,c) Photostriction produces a decrease of the area of the unit cell $A$ with $n_c$ and (b,d) a simultaneous decrease of unit cell area $A$ and polarization $P$.}\label{fig:fig3}
\end{figure}

Figures~\ref{fig:fig2}(a) and \ref{fig:fig2}(b) display a {\em decrease} of $a_1$ ($a_1(n_c>0)<a_{1,0}$) and an {\em increase} of $a_2$ ($a_2(n_c>0)>a_{2,0}$) for both SnS and SnSe monolayers. More specifically, the ratio $a_1(n_c)/a_2(n_c)$ is equal to $-0.58$ for the $nX$ transition and $-0.21$ at the $nY$ transition for SnS. In SnSe, $a_1(n_c)/a_2(n_c)=-0.26$ ($nX$) and $-0.16$ ($nY$). (For reference, Poisson ratios are 0.36 and 0.42 for SnS and SnSe, respectively \cite{liapl}.) In addition, a compression of the unit cell area $A$ {\em versus} $n_c$ ($A<A_0$) is illustrated in Figs.~\ref{fig:fig3}(a,c). Figs.~\ref{fig:fig2}(a-b) contain the first prediction of photostrictive effects in 2D materials; they  open a completely unexplored door for studies of {\em coupled mechano-opto-electronic effects} on these 2D compounds.

Furthermore, the rather large change on $a_1$ and $a_2$ in Figs.~\ref{fig:fig2}(a-b) ($|\Delta a_i/a_{i,0}|\sim 10^{-3}$) (under experimentally accessible photoexcited charge carrier densities $n_c\sim 10^{12}$/cm$^2$ \cite{Gedik}) is one to two orders of magnitude larger than that reported for bulk ferroelectrics and hence quite encouraging: such large values of $\Delta a_i/a_{i,0}$ place these new photostrictive 2D materials in a class of their own.

As seen in Figs.~\ref{fig:fig2}(c-d), the modification of $a_1$ and $a_2$ leads to a decrease of the electric dipole $P$ (obtained from Born effective charges) and the simultaneous decrease of $A$ and $P$ documented in Fig.~\ref{fig:fig3}(b,d).

\begin{figure}[tb]
\includegraphics[width=0.485\textwidth]{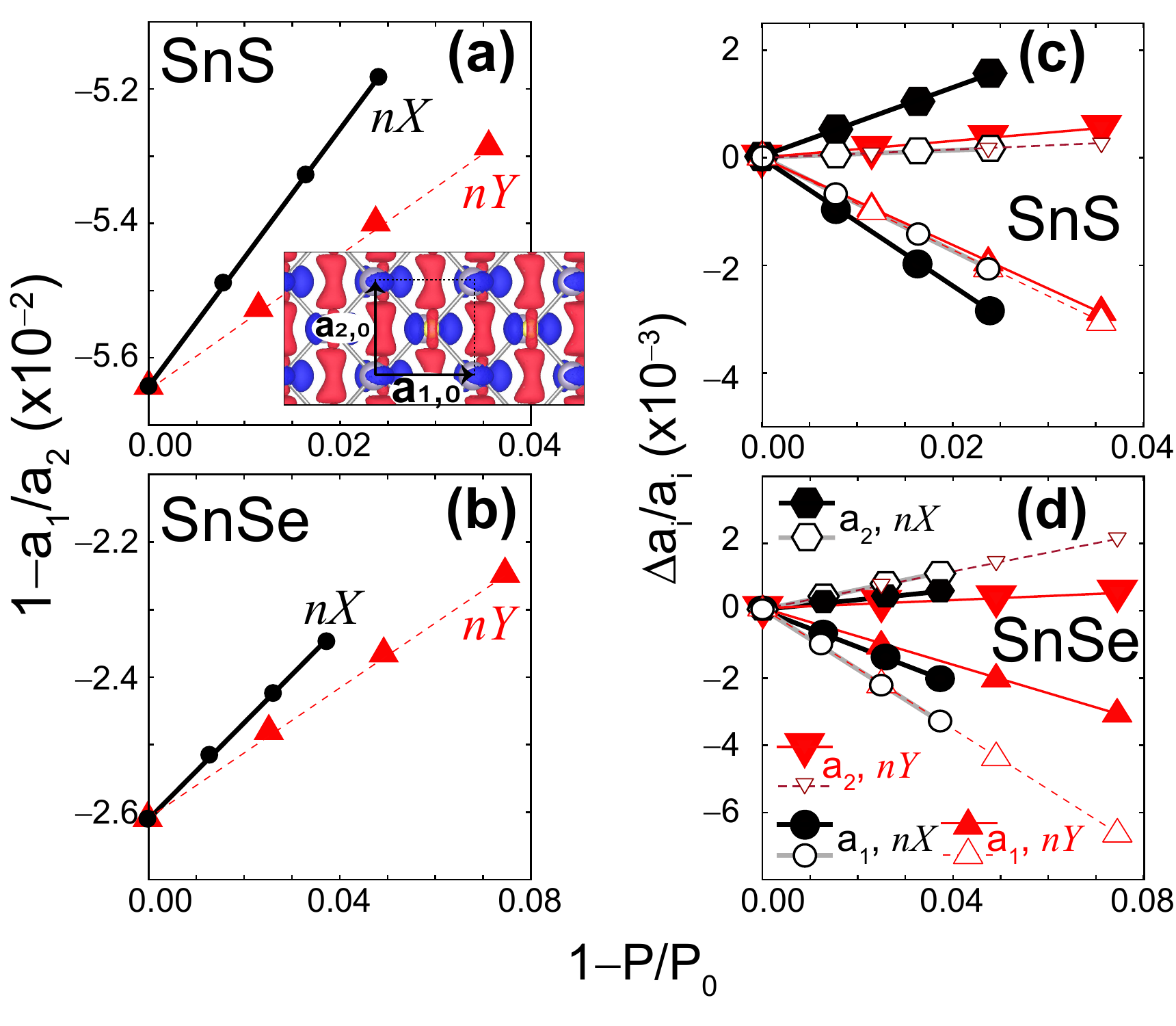}
\caption{(a-b) The decrease on dipole moment $P$ is linearly dependent on a decrease on the ratio $a1/a2$. (c-d) The decrease of lattice parameters with polarization $P$ seen with solid symbols for both materials and direct transitions ($nX$ or $nY$) arises from an inverse piezoelectric effect, as calculated from Eqn.~1 and shown in open symbols.}\label{fig:fig4}
\end{figure}

The reduction of $P$ seen in Figs.~\ref{fig:fig2}(c-d) is related to the anisotropic change in lattice constants seen in Fig.~\ref{fig:fig4}(a) and ~\ref{fig:fig4}(b) for SnS and SnSe, respectively. SnS and SnSe monolayers host an in-plane $P$ parallel to the $\mathbf{a}_1$ lattice vector that becomes reduced as the ratio $a_1/a_2$ approaches unity \cite{Mehboudi2016b}: this is why the polarization $P_0\equiv P(n_c=0)=2.77\times 10^{-10}$ C/m for SnS ($a_{1,0}/a_{2,0}$=1.056) is larger than that for SnSe ($P_0=2.06\times 10^{-10}$ C/m, and $a_{1,0}/a_{2,0}$=1.026) already and, within a given material, the reason for the thermally-induced ferroelectric-to-paraelectric transition for a sudden change of the structural order parameter $a_{1,0}(T)/a_{2,0}(T)$ towards unity without illumination, where $P_0(T_c)$ goes all the way to zero \cite{Mehboudi2016b,Yang2} at the transition temperature $T_c$. Photostriction is a new (optical) handle to tune the lattice parameters that reduces $a_1$ and increases $a_2$, regardless of valley edges being excited ($nX$ or $nY$). The coupling among polarization and structure, to be latter discussed, makes $P$ consequently smaller.

We showed the tunability of $a_1$ and $a_2$ with chemistry \cite{Mehboudi2016} and temperature before \cite{Mehboudi2016b}. In addition, a remarkable tunability under illumination has been unveiled now.

The inset in Fig.~\ref{fig:fig4}(a) displays the charge rearrangement upon illumination of the SnS monolayer as a charge density difference plot between the photoexcited state at k-point $nX$ and the ground state, prior to any structural optimization. Red isosurfaces (with a density of $+5\times 10^{5}$ electrons/\AA${^3}$) indicate excess electronic charge and blue isosurfaces (with a density of $-5\times 10^{5}$ electrons/\AA${^3}$) show electron depletion. As indicated earlier on, charge rearrangement is bound to occur upon photoexcitation, and regardless of the numerical method employed (i.e., that in Refs.~\cite{Hedin,Louie}, Ref.~\cite{Ullrich}, or the present one~\cite{Gorling,Martin,gonze2009abinit}, which permits a comparatively small time-consuming tracking of the structural distortion). Local exciton wavefunctions on GeS and GeSe shown in Ref.~\cite{excitons} will also necessarily perturb the initial electric dipole, and are bound to lead to a structural distortion akin to the one shown here. The point is that, although numerical estimates will naturally depend on method, the modification of the lattice structure with light is being successfully demonstrated for these compounds here.

Photoexcitation in these 2D ferroelectrics originates from an inverse piezoelectric effect \cite{Laurent} as follows. In 2D, the dielectric susceptibility $\chi_i^{2D}$ and the dielectric tensor $\epsilon_i$ (both diagonal) are related as $\chi_i^{2D}=\frac{(\epsilon_i-1)}{4\pi}L$ \cite{gomes2,r24,r25}, where $L$ is the vertical separation between (periodic) layers. This way, taking the observed change in polarization $P-P_0$ (which only occurs along the $x-$axis) and considering the $mm2$ point-symmetry of these compounds, lattice parameters $a_i$ ($i=1,2$) must evolve as \cite{gomes2,Nye}:
\begin{equation}
\frac{\Delta a_i}{a_{i,0}}=\frac{d_{i1}}{8\pi \epsilon_0 \chi_1^{2D}}(P-P_{0}),
\end{equation}
as represented by the open symbols in Figs.~\ref{fig:fig4}(c-d). $\chi_1^{2D}$ is taken as is from Ref.~\cite{gomes2} --and expressed in \AA{}-- and relaxed-ion values for $d_{i1}$ were taken from Ref.~\cite{liapl}; $\epsilon_0$ is the permittivity of vacuum \footnote{An additional factor of two multiplying $\chi$ arises from the length scale chosen for the vertical direction:  $a_3$ in Ref.~\cite{liapl}, and $a_3/2$ in Ref.~\cite{gomes2}.}. These predicted trends are of the same order of magnitude to the values of $a_1$ and $a_2$ determined upon a full optimization of the photoexcited structure, and they imply that photostriction is primarily produced by an inverse piezoelectric effect due to a dipole screening by the photoexcited charge carriers.

Note that the slope in Eqn.~1 is independent of the valley being photoexcited ($nX$ or $nY$), making the predicted values for $\Delta a_{i}/a_{i,0}$ lie upon the same straight line. While we assumed in the model the same Born effective charges for the two valleys in computing $P$, the actual polarization may be slightly different when exciting the $nX$ or the $nY$ valley. As shown next, an electronic/hole pressure may also produce slight differences in slope when exciting different valleys.

\begin{table}
 \caption{In-plane stress (in GPa) prior to structural relaxation arising from photoexcitation at the $nX$ and $nY$ points for $n_c(1)$. $\Delta a_i/a_{i,0}$ ($i=1,2$) below must be scaled by $10^{-4}$.}
\centering
\begin{tabular}{cccc||cccc}
\hline
SnS& & & &SnSe & & & \\
$nX$&$nX$ &$nY$&$nY$ &$nX$&$nX$ &$nY$&$nY$ \\
   \hline
$\sigma_{xx}$ & $\sigma_{yy}$ & $\sigma_{xx}$ & $\sigma_{yy}$&
$\sigma_{xx}$ & $\sigma_{yy}$ & $\sigma_{xx}$ & $\sigma_{yy}$\\
\hline

0.029&0.023&0.023&0.029&0.027&0.024&0.024&0.026\\
\hline
\hline
$\frac{\Delta a_1}{a_{1,0}}$ & $\frac{\Delta a_2}{a_{2,0}}$ & $\frac{\Delta a_1}{a_{1,0}}$ & $\frac{\Delta a_2}{a_{2,0}}$&
$\frac{\Delta a_1}{a_{1,0}}$ & $\frac{\Delta a_2}{a_{2,0}}$ & $\frac{\Delta a_1}{a_{1,0}}$ & $\frac{\Delta a_2}{a_{2,0}}$\\
\hline
$-8.3$ & 0.5 & $-4.6$ & $-1.9$ & $-7.9$ & $-0.3$ & $-6.1$ & $-1.3$\\
\hline
 \end{tabular}
\end{table}

Elongation of in-plane lattice vectors leads to positive stress. But when let to relax, the material {\em contracts} back to its original structure. In general, any structure with positive stress will contract in response. Therefore, in a first approximation, the lattice also displays an elastic {\em response} (having a {\em negative} sign) given by:
\begin{equation}
\frac{\Delta a_i}{a_{i,0}}=C^{-1}_{ij}(-\sigma_{j}).
\end{equation}
Using the elastic coefficients from Ref.~\cite{gomes2}, and the in-plane stress recorded in Table I for $n_c(1)$ from the initial photoexcited structure {\em prior to any structural relaxation}, we obtain changes of $\Delta a_i/a_{i0}$ from Eqn.~(2) that are an order of magnitude smaller than those seen in Fig.~\ref{fig:fig4}(c-d). This way, the numerical results from the structural optimization must be dominated by the inverse piezoelectric effect, thus showing the relevance of ferroelectricity for this effect to occur in 2D materials.

The trends in Figs.~\ref{fig:fig2}-\ref{fig:fig4} are similar to those for BiFeO$_3$, which implies  similar mechanisms at play. Experimental realization of ferroelectric 2D monochalcogenide monolayers \cite{Chang274} enhances the present relevance of this work, and brings optimism in that the unique effects here described will soon be experimentally verified.

In conclusion, the $\Delta$SCF method has been employed to predict photostriction of a novel family of 2D ferroelectrics known as group IV monochalcogenides. Photostriction results in a decrease of the larger lattice vector $a_1$ and an increase of the smaller one $a_2$, and it leads to a decrease of the area of the unit cell. It mainly arises from an inverse piezoelectric effect that reduces the dipole moment in the unit cell and contracts the lattice vector that is parallel to the electric dipole. The results provided in the present Letter continue to highlight unique properties of two-dimensional ferroelectrics and their potential usefulness for mechano-opto-electronic applications.

We are grateful to H. Churchill, B. Hamad and S. Sharifzadeh for discussions. R. Haleoot acknowledges funding from The Higher Committee For Education Development of Iraq. S. B.-L. was funded by an Early Career Grant from the US DOE (Grant DE-SC0016139). C.P. thanks the support from DARPA Grant No. HR0011-15-2-0038 (MATRIX program). B.X. and L.B. acknowledge the US AFOSR Grant No. FA9550-16-1-0065. Calculations were done at SDSC's {\em Comet} (XSEDE TG-PHY090002).


%

\end{document}